\documentstyle[draft,multicol,aps,psfig]{revtex}

\begin{document}

\draft

\title{
Two-finger selection theory in the Saffman-Taylor problem} 

\author{
F.X. Magdaleno and J. Casademunt}
\address{
Departament d'Estructura i Constituents de la Mat\`eria\\
Universitat de Barcelona,
Av. Diagonal, 647, E-08028-Barcelona, Spain
}

\maketitle

\begin{abstract}

We find that solvability theory selects a set of 
stationary solutions of the
Saffman-Taylor problem with coexistence of two \it unequal \rm 
fingers advancing with the same velocity but with different relative widths
$\lambda_1$ and $\lambda_2$ and different tip positions.
For vanishingly small dimensionless surface tension $d_0$, an infinite 
discrete set of values of the
total filling fraction
$\lambda = \lambda_1 + \lambda_2$ and of the relative individual 
finger width $p=\lambda_1/\lambda $ are selected
 out of a two-parameter continuous degeneracy. 
They scale as $\lambda-1/2 \sim d_0^{2/3}$ and $|p-1/2| \sim d_0^{1/3}$.
The selected values of $\lambda$ differ from those of the single finger case.
Explicit approximate expressions for both spectra are given.

\end{abstract}

\pacs{PACS numbers: 47.54.+r, 47.20.Ma, 47.20.Ky, 47.20.Hw}
\begin{multicols}{2}

\narrowtext

The Saffman-Taylor problem has played a central role in the field of
interfacial pattern selection in the last few decades\cite{reviews}.
It deals with the
morphological instability of the interface between two inmiscible fluids
confined in a quasi twodimensional (Hele-Shaw) cell, when the less viscous
fluid is displacing the more viscous one in a channel geometry\cite{Bensimon}. 
In particular, in their seminal work\cite{ST} 
Saffman and Taylor
called the attention upon the so-called selection
problem, namely the fact that
a unique finger-like steady state solution
is observed whereas a continuum of solutions is possible if surface
tension is neglected. Full analytical understanding of
the subtle role of surface tension acting
as the relevant selection mechanism was not achieved until much more
recently\cite{selection,Hong,selectionC,selectionS,Tanveer2}.
The resulting scenario of selection has been shown to apply with 
some genericity to 
other interfacial pattern forming systems, most remarkably in dendritic
growth\cite{reviews,Hong2}. 
On the other hand, despite
the relative analytical tractability of the problem,
the \it dynamics \rm of competing fingers is far
from being understood even at a qualitative level. 
Recently, it has been shown that in
general surface tension may affect the long time dynamics in an essential
way\cite{Tanveer1}. In the case of the dynamics of finger arrays, 
the effect of surface tension becomes particularly dramatic, 
showing that the qualitative picture of finger competition
based solely on the concept of screening of the laplacian field or the 
global instability\cite{Kessler} of a periodic finger array  
is insufficient\cite{Magdaleno1}.

Existence of multifinger stationary solutions of the zero surface tension 
problem has been known for a long time. In Ref.
\cite{Magdaleno1} it has been emphasized that multifinger stationary 
solutions are relevant to the issue of the dynamical role of 
surface tension. In particular the equal-finger fixed point has been 
pointed out as the relevant saddle-point to describe competition dynamics.
In connection with the phase flow structure around this fixed point, the 
problem of existence of unequal-finger fixed points with 
\it nonzero \rm surface tension has been posed\cite{Magdaleno1}.
Here we will extend selection theory 
to search for such solutions in the case of two fingers.
We will follow the formulation of Hong and Langer\cite{Hong}, which
is based on
a Fredholm solvability analysis of a non-selfadjoint problem 
defined through linearization about the zero surface tension solution,
together with WKB and steepest descent techniques. 
Despite the admitted objections to the full quantitative
validity of the approximations involved in this method
\cite{selectionC,Tanveer2}, 
it has been shown that it
leads to the correct qualitative picture of selection and the correct
scaling of solutions\cite{Tanveer2}. 
It is therefore suitable, for simplicity of calculus 
and presentation, for a first exploration
of novel situations such as the present one. 

Our starting point is the dynamical equation for the conformal mapping 
$f(w,t)$ which maps the interior of the unit circle in the complex plane 
$w$ into the viscous fluid region, 
with the unit circle $w=e^{i\phi}$ being mapped 
into the interface. 
Without loss of generality, we will assume a channel 
width $W=2\pi$ in the $y$ direction (with periodic boundary conditions)
and a velocity $U_{\infty}=1$
of the fluid at infinity. We define the velocity of the stationary solutions
of the interface as $U=\frac{1}{\lambda}$ where $\lambda$ is the total filling 
fraction of the channel by the invading fluid. 
The cartesian coordinates in the frame moving with velocity $U$ are given by
$z=x+iy=f(w,t) - Ut$.
The mapping $f(w,t)$ contains a logarithmic 
singularity which is due to the fact that we are mapping an unbounded 
domain (the semiinfinite strip) into 
the unit circle, in such a way that $f(w,t)+\ln w$ is always an analytic 
funtion in the interior of the unit circle.
The exact dynamical equation for the mapping can be written in the form
\begin{equation}
\label{eq:eqmotion}
Re(i\partial_{\phi}{f(\phi,t)}\partial_{t}{f^*(\phi,t)})=1 -
Ud_{0}\partial_{\phi} H_{\phi}[\kappa]
\end{equation}
which can be easily derived for instance from Ref.\cite{Bensimon}. 
Here $d_{0}$ is a dimensionless surface tension defined as 
$ d_0=\frac{\sigma b^{2}}{12 \mu U}$,
where $\sigma$ is the surface tension, b is the gap and $\mu$ is the 
viscosity. The curvature
$\kappa$ can be expressed in terms of $x(\phi)$ and $y(\phi)$ as 
\begin{equation}
\kappa(\phi)=
\frac{\partial_{\phi}^{2}{x}\partial_{\phi}{y}-
\partial_{\phi}^{2}{y}\partial_{\phi}{x}}{[(\partial_{\phi}{y})^2
+(\partial_{\phi}{x})^2]^{\frac{3}{2}}}.
\end{equation}
$H$ is a linear integral operator (Hilbert transform) 
which acts on a real $2\pi$-periodic function $g(\phi)$ according to
the definition 
\begin{equation}
\label{eq:hilbert}
H_\phi [g] = \frac{1}{2\pi} P \int_0^{2\pi} g(s) {\rm cotg} \frac{1}{2}(\phi-s)  ds.
\end{equation}
It follows from Eq.(\ref{eq:hilbert}) that 
the function $A(w)$ defined by $A( e^{i\phi}) = 
g(\phi) + i H_\phi [g]$ is  
analytic in the interior of the unit circle, and $Im(A(0))=0$. 

In the steady state we will have $\partial_{t}{f^*(\phi,t)}=U$, and 
Eq.(\ref{eq:eqmotion}) will read
\begin{equation}
\label{eq:steady}
-U\frac{dy}{d\phi}=1-Ud_{0}\frac{dH_{\phi}[\kappa]}{d\phi}
\end{equation}
We search for solutions of the generic type described in Fig.1. The total 
filling fraction $\lambda$ is splitted into two contributions $\lambda_1+
\lambda_2=\lambda$, and we define as a new selection parameter the 
relative finger width $p=\lambda_1/\lambda$. 
For simplicity we will consider fingers which are axisymmetric 
and for convenience we will fix the tip positions at
$\phi=\pi/2, 3\pi/2$, for all $\lambda$ and $p$. The filling fraction 
$\lambda$ ranges from $0$ to $1$. We  
take $\lambda_2 \leq \lambda_1$ so that $p$ ranges from $1/2$ to $1$.
In these conditions the two fingers  
correspond to the intervals $\phi_1=\frac{\pi}{2}(1-2p)$ to 
$\phi_2=\frac{\pi}{2}(1+2p)$ 
and $\phi_2$ to $\phi_3=2\pi+\phi_1$ respectively.

After integration over $\phi$ of Eq.(\ref{eq:steady}) we obtain
\begin{equation}
\label{eq:piecewise}
-Uy(\phi)=\phi-Ud_{0}H_{\phi}[\kappa]+c(\phi)
\end{equation}
where $c(\phi)$ is a piecewise constant function. The values it takes 
at the intervals  $(\phi_1,\phi_2)$ and  $(\phi_2,\phi_3)$ differ by 
the finite amount $\pi(U-1)$ which accounts for the discontinuity of 
$y$ and the finite flux at the points $\phi_1, \phi_2$.

After Hilbert transform of Eq.(\ref{eq:piecewise}) and 
using that 
$H_\phi^2[g] = -g(\phi)$, we obtain
\begin{eqnarray}
\label{eq:hiltrans1}
-d_{0}\kappa(\phi)+x(\phi)=x_{0}(\phi)\\
\label{eq:hiltrans2}
y(\phi)+\phi=const +H_{\phi}[x].
\end{eqnarray}
where Eq.(\ref{eq:hiltrans2}) is just an expression of analyticity of 
$f(w,t)+\ln w$. The function $x_0(\phi)$ is found explicitely as 
$x_0(\phi)=H_\phi[g]$ with $g(\phi)=(\lambda-1)\phi+c(\phi)$, and by 
construction it corresponds to the solution of the zero surface tension 
case. In our case it reads
\begin{eqnarray}
x_{0}(\phi)=(1-\lambda) \ln (2|\sin\phi-\cos p\pi|).
\end{eqnarray}

Completed with $y_0(\phi)=-\lambda(\phi+c(\phi))$, this gives a two
parameter class of exact solutions of the type of Fig.1, for $d_0=0$.
Both $\lambda$ and $p$ can be varied continuously within their natural
range. The difference $\Delta_x$ between the x-coordinate of the two tips is 
given by 
\begin{equation}
\label{eq:delta}
\Delta_{x} = (1-\lambda)\ln \frac{1-\cos p\pi}{1+\cos p\pi}.
\end{equation}
These solutions are precisely those studied in Ref.\cite{Magdaleno1}.

The present formulation has some interesting advantages
over the traditional approach of McLean and Saffman\cite{selection}
 for instance in that
the zeroth order solution
is obtained naturally as an explicit outcome of the method and that it 
is more amenable to generalization, for instance to a larger number of
fingers\cite{Magdaleno2}.

We now proceed by assuming
$x(\phi)=x_{0}(\phi)+d_{0}x_{1}(\phi)$ and linearizing on $x_{1}(\phi)$ 
but keeping all singular terms necessary for selection. (Nonlinear effects 
are expected to introduce only a slight quantitative correction to the final 
spectrum of selection\cite{selectionC,Tanveer2}.) 
Using the relation $y_{1}(\phi)=H_{\phi}[x_{1}]$
we get
\begin{eqnarray}
\label{eq:integro}  
d_{0}\frac{d^{2}x_{1}}{d\phi^2}+d_{0}p(\phi)\frac{d^{2}H_{\phi}[x_{1}]}
{d\phi^2}+r(\phi)x_{1}=\mu(\phi)
\end{eqnarray}
where  $r(\phi)$ and $p(\phi)$ are given by 
\begin{eqnarray}
r(\phi)=
\lambda^{2} \frac{|q(\phi)|}{q^4(\phi)}
[(q(\phi))^2 +\frac{1}{\beta^2}\cos^{2}\phi ]^\frac{3}{2}\\
p(\phi)=
\frac{1}{\beta}\frac{\cos\phi}{q(\phi)}
\end{eqnarray}
with $q(\phi)=\sin\phi -\cos p\pi$ and $\beta=\frac{\lambda}{1-\lambda}$.
Explicit knowledge of $\mu (\phi)$ is not necessary for the solvability analysis.
First order derivatives are subdominant as $d_0 \rightarrow 0$ and 
have been omitted\cite{Hong}.

The linear operator on the lhs of Eq.(\ref{eq:integro}) can be seen as a
$2\times 2$ matrix operator acting on a vector of two components
$x_1^+(\phi)$ and $x_1^-(\phi)$ which are defined respectively on the intervals
$(\phi_1,\phi_2)$ and $(\phi_1,\phi_3)$. Inserting an ansatz of WKB form 
with a point of stationary phase of the solution in the upper (or lower) 
complex plane\cite{Hong} one can show, using steepest descent techniques, 
that the off-diagonal terms of Eq.(\ref{eq:integro}) lead to 
exponentially small contributions. 
As a consequence, to leading order the
problem is decoupled into two separate problems defined in two disjoined
intervals.
Similarly, neglecting exponentially small terms, the integral part of the 
diagonal terms takes a purely differential form in the complex 
plane\cite{Hong}. The change 
of variables $\eta=-\frac{1}{\beta} \frac{\cos\phi}{\sin\phi-\cos p\pi}$ 
maps separately each of the two disjoint intervals above into the 
whole real axis $\eta \; \epsilon \; (-\infty,\infty)$. 
Therefore, to leading order we endup with two (complex) 
differential equations of the form
\begin{eqnarray}
d_{0}\frac{d^{2}x_1^{\pm}}{d\eta^2}+ Q_{\pm}(\eta)x_1^{\pm}=R_{\pm}(\eta).
\end{eqnarray}
which are mutually independent but linked through the dependence on the 
parameters $\lambda$ and $p$.
More details of this derivation will be presented elsewhere\cite{Magdaleno2}.
We define two
solvability functions as
\begin{eqnarray}
\Lambda_{\pm}(\lambda,p;d_{0})=
\int_{-\infty}^{\infty} \tilde{x}^{\pm}(\eta) R^{*}_{\pm}(\eta)
\,{\rm d}\eta
\end{eqnarray}
where $\tilde{x}^{\pm}(\eta)$ are eigenfunctions of the null space of the
adjoint operators of the respective homogeneous equations\cite{Magdaleno2}.
To enforce solvability we now have to impose the simultaneous vanishing 
of the two solvability functions $\Lambda_{\pm}(\lambda,p;d_{0})=0$.
These two conditions will fix the discrete spectra of possible values of
both $\lambda$ and $p$.

Within the WKB approximation, the two solvability functions take the form
\begin{eqnarray}
\Lambda_{\pm}(\lambda,p;d_{0})=
\int_{-\infty}^{\infty} G_{\pm}(\eta) e^{\frac{1}{\sqrt{d_{0}}}
\Psi_{\pm}(\eta)} \,{\rm d}\eta
\end{eqnarray}
where
\begin{eqnarray}
\Psi_{\pm}(\eta)= i\lambda\beta
\int_{0}^{\eta}
\frac{(1-i\eta')^{\frac{1}{4}}(1+i\eta')^{\frac{3}{4}}}{1+\beta^{2}\eta'^{2}}
\nonumber\\
\times \left( 1 \mp \frac{\cos p\pi}{\sqrt{1+ \eta'^{2}
\beta^{2}\sin^{2}p\pi }}
\right)
\,{\rm d}\eta'.
\end{eqnarray}
In order to 
estimate the solvability 
functions in the steepest descent approximation,  only the 
form of $\Psi_{\pm}(\eta)$ is required. 
The singularity structure of $\Psi_{\pm}(\eta)$ is such that the cases 
$p=1/2$ and $p\neq 1/2$ must be treated separately. 
The first case (two identical 
fingers) degenerates 
into the usual single finger problem. 
For $p>\frac{1}{2}$, a more complicated 
singularity structure is revealed. In the upper half complex plane of $\eta$, 
we find that 
$\frac{d\Psi_{+}(\eta)}{d\eta}$ has a new branch point at 
$\eta=i/\beta \sin p\pi$,
 in addition to the singularities that 
were present in the single finger problem, namely, a branch point at 
$\eta=i$ and a pole at $i/\beta$. On the other hand, 
$\frac{d\Psi_{-}(\eta)}{d\eta}$ has the branch point at $\eta=i$ and 
the new one at 
$\eta=i/\beta \sin p\pi$,
whereas the pole at
$i/\beta$ is suppressed.
Since $1/\beta \sin p\pi > 
1/\beta$, we obtain that $\beta > 1$  
is a necessary condition for the first solvability function 
$\Lambda_{+}(\lambda,p;d_{0})$ to oscillate, and therefore generate zeroes.
 We thus recover the condition 
$\lambda > 1/2$ of the single finger case, but 
now for the total filling fraction. The equivalent condition for 
$\Lambda_{-}(\lambda,p;d_{0})$ is $\beta \sin p\pi > 1$ so that the new
singularity at $\eta=i/\beta \sin p\pi$ stands below $\eta=i$. This condition 
also implies that in the contour integration for $\Lambda_{+}(\lambda,p;d_{0})$
we will always pick up a contribution from this new singularity.

By deforming the contour integral as indicated in Fig.2, and following 
Ref. \cite{Hong} in identifying the crossover from oscillating 
to non oscillating behaviour of the solvability functions, we  
obtain the scaling of both $\lambda$ and $p$ with $d_0$ to be 
$(\lambda-\frac{1}{2})
\sim d_{0}^{2/3}$ and $|p-\frac{1}{2}| \sim d_{0}^{1/3}$.
According to Eq.(\ref{eq:delta}), the resulting scaling for the tip 
difference is $\Delta_x \sim d_{0}^{1/3}$.

An explicit (approximate) discrete spectra of selected 
values of $\lambda$ and $p$ for small $d_0$ will be
given by the condition 
$\cos (\frac{\Psi_{\pm}(i+0)-\Psi_{\pm}(i-0)}{2i\sqrt{d_{0}}})=0$.

From the condition $\Lambda_{-}(\lambda,p;d_0)=0$ we thus obtain
\begin{eqnarray}
\label{eq:discret2}
\frac{1}{\sqrt{d_{0}}}\frac{(1-\lambda)^2}{\lambda} 
I(\beta,p) = m-\frac{1}{2}  
\end{eqnarray}
with $m=1,2,...$ and where
\begin{eqnarray}
I(\beta,p)= -\frac{1}{2\pi} {\rm cotg} p\pi \int_{0}^{u_{1}}
\frac{u^{\frac{3}{4}}  H(u;\beta,p)   }{(u_{3}-u)
(u_{1}-u)^{\frac{1}{2}}} \,{\rm d}u 
\end{eqnarray}
with the regular part of the integrand 
$H(u;\beta,p)=(2-u)^{\frac{1}{4}} (u_{4}-u)^{-1}
(u_{2}-u)^{-\frac{1}{2}}$
and $u_{1,2}=1 \mp 1/\beta \sin p\pi$, $u_{3,4}=1 \mp 1/\beta$.

Finally, from condition $\Lambda_{+}(\lambda,p;d_0)=0$, expressing 
Eq.(\ref{eq:discret2}) to leading order and using properties of hypergeometric
functions, the two selection conditions 
can be combined to read\cite{Magdaleno2} 
\begin{eqnarray}
\label{eq:discret1}
\frac{1}{\sqrt{d_{0}}}(2 \lambda-1)^{\frac{3}{4}}=n \\
\label{eq:discret3}
S(\alpha)
=\frac{1}{n} (m-\frac{1}{2})
\end{eqnarray}
with $n=1,2,...$ and where 
\begin{equation}
S(\alpha) = \frac{3\sqrt{2\pi}}{5\Gamma^{2}(\frac{1}{4})}
\; _{2}F_{1}
\left(\frac{5}{4},\frac{1}{2};\frac{9}{4};1-\alpha \right)
(1-\alpha)^{\frac{5}{4}}.
\end{equation}
$_{2}F_{1}$ is a hypergeometric function 
\cite{abramowitz} and 
$\alpha=\frac{\pi^{2}}{4} \frac{(p-\frac{1}{2})^{2}}
{2 \lambda-1} $ is of order $(d_0)^0$ and ranges from $0$ 
to $1$.

Eq.(\ref{eq:discret1}) determines a set of discrete 
values of $\lambda$. 
Notice that these are given independently of $p$ 
but the set of values are inserted between those
of the single finger case ($p=1/2$), which in the same approximation 
are given by $\frac{1}{\sqrt{d_{0}}}(2 \lambda-1)^{\frac{3}{4}}
= n - \frac{1}{2}$ in place of Eq.(\ref{eq:discret1}).
On the other hand, 
the lhs of Eq.(\ref{eq:discret3}) is a monotonically decreasing function 
of $\alpha$ which varies continuously from $1/4$ (at $\alpha=0$) to $0$
(at $\alpha=1$)\cite{abramowitz}. 
Solving Eq.(\ref{eq:discret3}) for $\alpha$ produces solutions
with $p \neq 1/2$. These will exist whenever
$\frac{1}{n} (m-\frac{1}{2}) < \frac{1}{4}$. For a given $n$, the solutions 
are labeled by $m=1,2,...$ up to the integer part of 
$(n+1)/4$. Therefore, the first solution with $p \neq 1/2$ will appear 
at $n=3$ and gives $|p-\frac{1}{2}| \simeq 0.3886\; d_0^{1/3}+...$
For fixed 
$m$, $p$ is an increasing function of $n$ (like $\lambda$), but 
for fixed $n$, $p$ has its maximum value at $m=1$ and then decreases 
with $m$ (See Fig.3). 
The spectra here derived must be taken with some caution, since they are only 
approximate.
An exact calculation to lowest order in 
$d_0$ should include nonlinear effects and a proper treatment of 
the turning points in the WKB analysis, but the corrections are
expected to be quantitatively small\cite{footnote}. 
More details will be presented elsewhere\cite{Magdaleno2}.


Concerning the stability of these solutions, 
it is reasonable to presume that, in general,
they will be globally unstable, such as established numerically
for the equal finger array ($p=1/2$) in Ref.\cite{Kessler}.
This implies that they would only be directly observable as a transient 
slowing down of the competition dynamics 
whenever an initial condition is prepared close to any of those 
solutions. 
From a dynamical systems point of view,
we would like to emphasize that 
the knowledge of the fixed points, 
even if unstable,
is always relevant to elucidate the topological features 
of the phase space flow, and therefore to gain insight and 
qualitative understanding 
of the dynamics. In particular the location of these new fixed points 
will definitely affect the path in phase space describing the 
transient dynamics from a nearly equal finger array towards the 
single finger attractor. 
This point of view was developed in Ref.\cite{Magdaleno1} 
to study the dynamical role of surface tension. In that spirit it was  
pointed out that a 
generalized solvability scenario of selection\cite{reviews,selection} 
could hold to some extent 
for the dynamics\cite{Magdaleno1,Magdaleno2}. 


We conclude by remarking that, although the present solvability analysis 
is not a rigorous proof of existence of solutions 
it reveals by itself
a quite unexpected richness of the problem. 
It would be interesting to search for
these solutions numerically or by other more rigorous means
\cite{selectionC,Tanveer2}.
The sole existence of the predicted solutions and its presumable generalization 
to a larger number of fingers has important consequences on the
physical picture of finger competition, 
which turns out to be much more complex than common arguments 
of laplacian screening seem to suggest. The common picture, according to which 
fingers slightly ahead escape from their neighbors, is not 
necessarily valid in general because of the existence of growth modes 
with unequal \it noncompetiting \rm fingers. 
For vanishingly small surface tension, 
however, these modes collapse and only the equal-finger multifinger mode 
($p=1/2$) survives as a stationary state.
Finally, given the genericity of the solvability 
mechanism of selection, this opens the possibility of finding similar
solutions in related problems such as needle crystal growth or 
viscous fingering in circular geometry.  

We acknowledge financial support from the Direcci\'on General de
Ense\~{n}anza Superior (Spain) under Project
PB96-1001-C02-02. 
F.X.M. also acknowledges financial support from the Comissionat
per a Universitats i Recerca (Generalitat de Catalunya).

\ \\

Fig. 1. Typical configuration of a two-finger stationary solution.

Fig. 2. Deformation of the steepest descent contour of integration in the
complex $\eta$ plane with $\beta>1$  and $p>\frac{1}{2}$ (a) for
$\Lambda_{+}$; (b) for $\Lambda_{-}$.

Fig. 3. Spectrum of $p$ as a function of $n$ for different values of $m$.

\end{multicols}


\begin{references}
\bibitem{footnote}
The proper WKB treatment of the turning points by matching of the inner 
and outer problems has been carried out up to two numerical constants, 
which appear on the rhs of Eqs.(\ref{eq:discret1}) and (\ref{eq:discret3}). 
The general picture is thus only slightly modified quantitatively.
More details will be presented elsewhere\cite{Magdaleno2}.
\end{references}
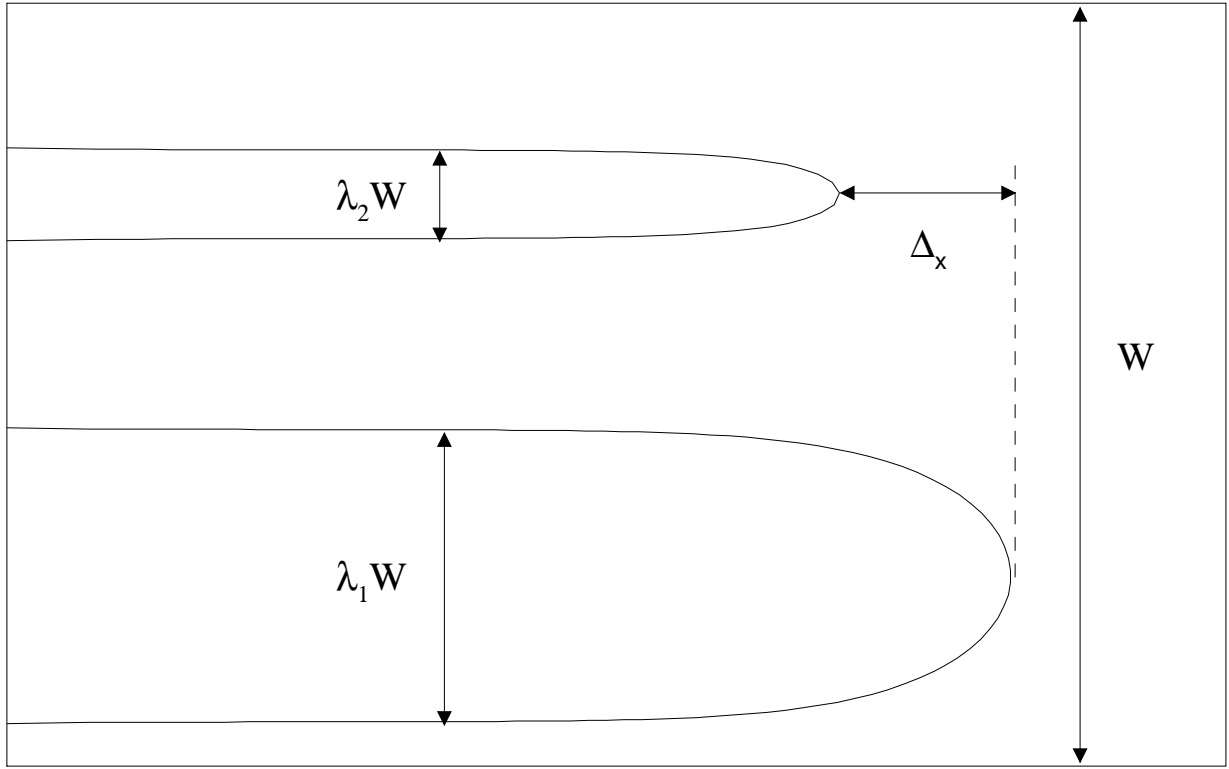
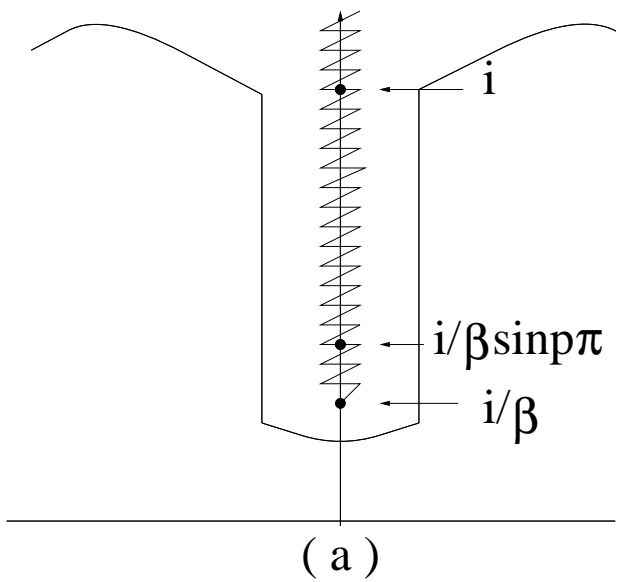
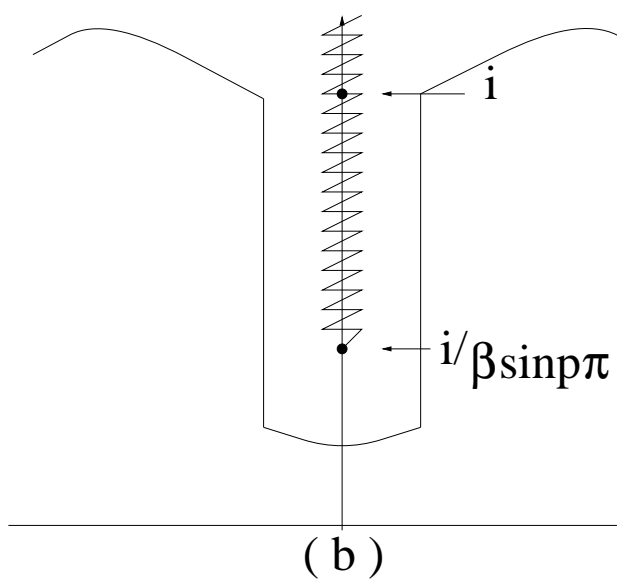
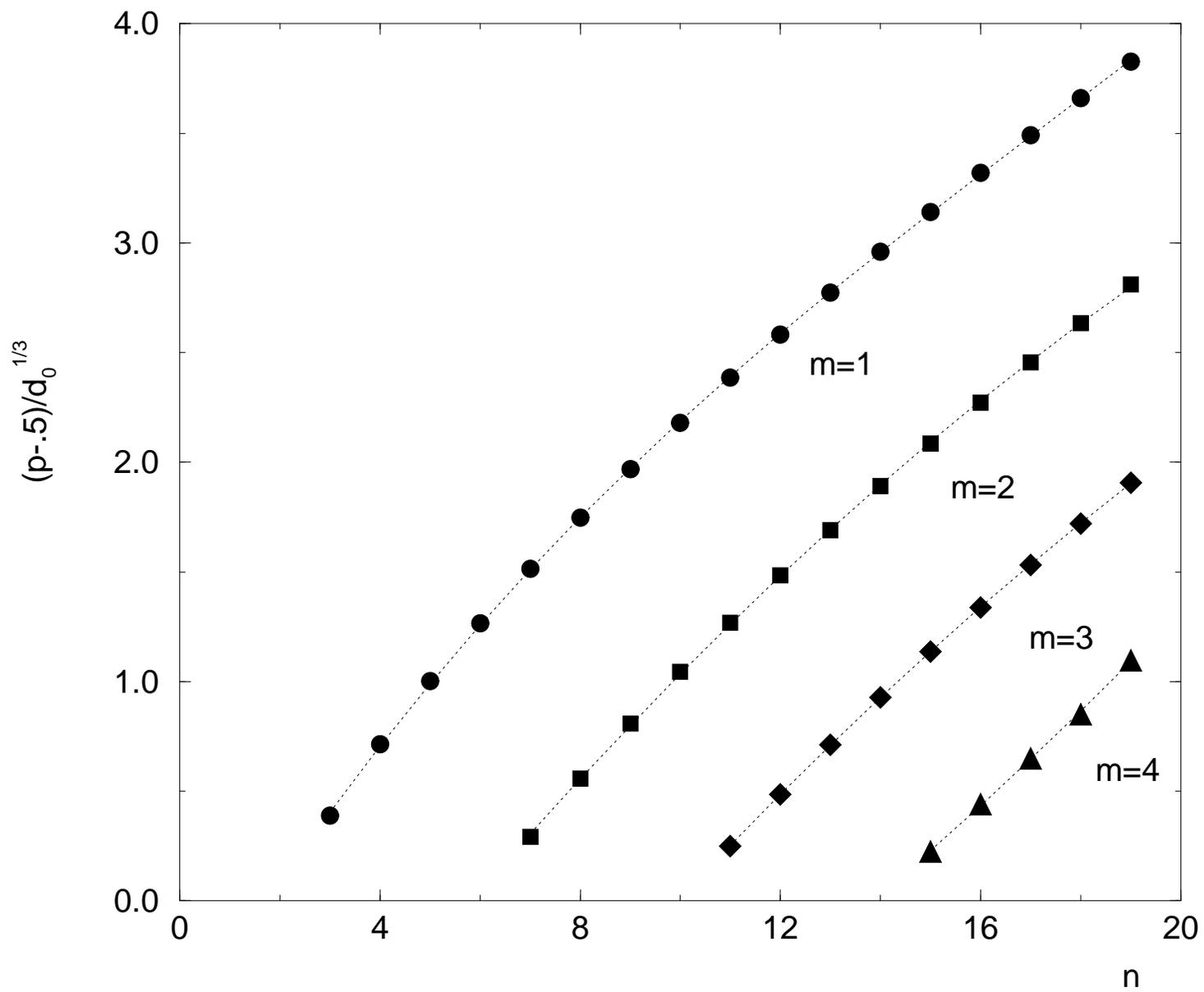
\end{document}